\documentclass[journal]{IEEEtran}

\usepackage[cmex10]{amsmath}
\usepackage{bm}
\usepackage{amsfonts}
\usepackage{amssymb}
\usepackage{relsize}
\usepackage{cleveref}
\usepackage{cuted}
\usepackage[caption=false]{subfig}
\usepackage{graphicx}
\usepackage{algorithm}
\usepackage{comment}
\usepackage[keeplastbox]{flushend}

\allowdisplaybreaks
\DeclareMathOperator{\sinc}{sinc}

\begin{document}

\title{Modeling Transmission and Radiation Effects when Exploiting Power Line Networks for Communication }

\author{\IEEEauthorblockN{Davide Righini, Federico Passerini, and Andrea M. Tonello}
\thanks{Davide Righini, Federico Passerini and Andrea M. Tonello are with the Ecosys Lab, University of Klagenfurt, Austria. 
Email: {\{davide.righini, federico.passerini, andrea.tonello\}@aau.at}.
\newline \noindent
A version of this article has been accepted for publication on IEEE Transactions on Electromagnetic Compatibility, Special Issue on ``Advances in Modeling, Measurement and Design of Discontinuities and Their EMC and SI/PI Effects on Wired Communication Links''.

\noindent IEEE Copyright Notice: \textsf{\copyright} 2017 IEEE. Personal use of this material is permitted. Permission from IEEE must be obtained for all other uses, in any current or future media, including reprinting/republishing this material for advertising or promotional purposes, creating new collective works, for resale or redistribution to servers or lists, or reuse of any copyrighted component of this work in other works.}}

\maketitle

\begin{abstract}
Power distribution grids are exploited by Power Line Communication (PLC) technology to convey high frequency data signals.  The natural conformation of such power line networks causes a relevant part of the high frequency signals traveling through them to be radiated instead of being conducted. This causes not only electromagnetic interference (EMI) with devices positioned next to power line cables, but also a consistent deterioration of the signal integrity. Since existing PLC channel models do not take into account losses due to radiation phenomena, this paper responds to the need of developing  accurate network simulators. A thorough analysis is herein presented about the conducted and radiated effects on the signal integrity, digging into differential mode to common mode signal conversion due to network imbalances. The outcome of this work allows each network element to be described by a mixed-mode transmission matrix. Furthermore, the classical per-unit-length equivalent circuit of transmission lines is extended to incorporate radiation resistances. The results of this paper lay the foundations for future developments of comprehensive power line network models that incorporate conducted and radiated phenomena.
\end{abstract}


\IEEEpeerreviewmaketitle

\section{Introduction}

\IEEEPARstart{P}{ower} line communication (PLC) is nowadays a widespread and compelling technology to convey information both in the context of Smart Grids and In-Home environments \cite{lampe2016power}. Its intrinsic property of using the existing cable infrastructure has the advantage for utilities of being proprietary and for residential users of granting wider range and possibly higher coverage compared to WiFi technologies.
On the other side, power lines are made of metallic cables that inherently radiate power when excited with a current signal, i.e. whenever a PLC signal is transmitted. Although this phenomenon is known by the PLC community since its foundation, dealing with it has always been challenging \cite[Part II]{Berger:2014:MPL:2636789}. Among the papers about this topic, \cite{kikkertradiation} presents a characterization of the radiation pattern in simple distribution lines, while \cite{paganireversal} presents a signal processing technique to reduce the radiated emission. However, to date a thorough characterization of radiation phenomena in power line networks is still missing. A proof of this is the fact that PLC standards fix transmission power limits based on empirical procedures \cite{7867791}. 

In this paper, we want to propose a comprehensive model of power line networks (PLNs) based on the physical description of both the conducted and radiated phenomena, with particular emphasis on broad band PLC that uses a frequency spectrum in the range 2-86 MHz and is applied in complex topologies as those found in home PLC networks. Such a model enables the simulation of complex networks so that PLC algorithms that cope with both transmission and radiation losses can be tested. Eventually, it can be used to analyze the interaction of PLC with wireless communication technologies.

In order to characterize the conducted and radiated phenomena that occur along every branch of a PLN, we make use of the transmission line super theory (TLST) \cite{1325788}. In particular, we first consider the case of a two-wire transmission line (TL) surrounded by an homogeneous dielectric and extend the work of \cite{ianconescu} including both differential and common mode signaling. We derive an analytic expression of the radiated power and propose an equivalent per-unit-length (PUL) model of the cable including the radiation losses, which we name multiple transmission and radiation line (MTRL) model. This model is applicable to beyond two conductors PLC networks. A scalable representation of a PLN can thereafter be made relying on the microwave network analysis theory \cite{Pozar}. We show that an equivalent mixed-mode transmission matrix \cite{4287166} can be derived from the MTRL model of each branch. Moreover, all the devices that cannot be physically modeled due to their complex geometrical and circuit structure (couplers, sockets, etc\dots) can be still represented by an equivalent mixed-mode transmission matrix (MMTM). The chain rule of transmission matrices finally allows the cumulative analysis of the network properties. Therefore, the herein proposed approach allows the development of a bottom-up PLC channel model that accounts for both conducted and radiated fields, which significantly extends  state-of-the-art PLC channel models presented in the literature \cite{versolatto2011an}.

This paper is organized as follows. In \Cref{sec:radiation}, we derive the radiation model for a two-wire transmission line and introduce the MTRL. In \Cref{sec:network_discontinuities}, we explain how mode conversion is caused by the discontinuities in power line networks, and how this affects both the radiation and the communication. \Cref{sec:network_model} is dedicated to the explanation of the mixed-mode transmission theory. Its application to the representation of single devices and the extension to power line networks is presented in \Cref{sec:application}. Measurements results are reported in \Cref{sec:comparison_simulation} and conclusions follow in \Cref{sec:conclusions}.

\section{Radiation losses in Power Lines}
\label{sec:radiation}

The classical TL theory is based on the telegraph model, which treats
TL as a series of RLGC cells. This model is suitable to describe conducted
phenomena in wired networks, since it provides a tool to analyze
the propagation of the voltage and current waves and to account for
dielectric and ohmic losses. However, the model does not take into
account the losses due to the propagation of radiated power in the
dielectric surrounding the TL. A recent work \cite{ianconescu} showed
that finite length TLs excited by differential mode currents radiate
power next to the line terminations. Our aim in this section is to
summarize the main results of the mentioned work and to extend the
results to TLs excited by common mode currents.

Considering a TL of length L made by two conductors in free space
and neglecting any dielectric loss for simplicity, the magnetic potential
vector $\mathbf{A}$ is \cite{hoop} 
\begin{multline}
\mathbf{A}\left(d\right)=\mu_{0}e^{-jkz}\intop_{0}^{L}dz'\ointop dcK_{f}\left(c\right)e^{-jkz'}G\left(d\right)\hat{\mathbf{z}}+\\
+\mu_{0}e^{jkz}\intop_{0}^{L}dz'\ointop dcK_{r}\left(c\right)e^{jkz'}G\left(d\right)\hat{\mathbf{z}}\label{eq:A}
\end{multline}
where $\mu_{0}$ is the vacuum permeability, $k$ is the propagation
constant, $K_{f}$ and $K_{r}$ are the forward and reverse surface
current waves respectively as function of the contour $c$, $G$ is
the Green's function
\begin{equation}
G\left(s\right)=\frac{e^{-jks}}{4\pi s}
\end{equation}
and $d$ is the distance of the observation point $\left(x,y,z\right)$
from the integration point $\left(x',y',z'\right)$. 

The solution of (\ref{eq:A}) for differential mode (DM) transmission
has been reported in \cite{ianconescu}, where the authors show that
the DM radiated power of a finite length TL is 
\begin{equation}
P_{rad}^{DM}=\frac{\eta_{0}\left(ka\right)^{2}}{2\pi}\left(\left|I_{f}^{DM}\right|^{2}+\left|I_{b}^{DM}\right|^{2}\right)\left(1-\sinc\left(4kL\right)\right),\label{eq:praddm}
\end{equation}
where $a$ is the distance between the conductors, $I_f$ and $I_b$ are the forward and backward traveling currents respectively.
The paper demonstrates that a TL excited by a DM radiates only from
a region around its terminations. To confirm this, (\ref{eq:praddm})
shows that if $L$ is much greater than the current wavelength $\lambda$, $P_{rad}^{DM}$ tends to a constant
value, which means that the central part of a long cable is not a source
of significant radiation. Conversely, $P_{rad}^{DM}$ is negligible
when $L\ll\lambda$, rapidly increases and can be already considered
stable at $L\simeq\lambda$. Given the fact that the cable radiates
symmetrically from the terminations, a model for the PUL radiation
resistance $R^{DM}(l)$ can be found by considering how half of $P_{rad}^{DM}$
changes up to the mid point of the line, which results in 
\begin{multline}
R^{DM}(l)=\frac{P_{rad}^{DM}}{2\left(\left|I_{f}^{DM}\right|^{2}+\left|I_{b}^{DM}\right|^{2}\right)}=\\=\frac{\eta_{0}\left(ka\right)^{2}}{4\pi l}\left(\sinc\left(4kl\right)-\cos\left(4kl\right)\right),\label{eq:rdm}
\end{multline}
which is function of the distance $l$ from the termination point. 

For what concerns the common mode (CM) transmission, a reference plane has to be taken into account in the system to let the CM current circulate. It can be noticed that for $d\gg L,\lambda$ the two lines are equivalent to a single wire which radiates double the power of a single wire alone. The radiated power can be obtained by solving \eqref{eq:A} for the single wire in free space, considering the array factor given by the image theory \cite{Balanis:2005:ATA:1208379}.  
However, if $L\lesssim\lambda$, the equivalent wire behaves like a long wire antenna, which is a kind of slow wave traveling antenna and whose properties have been thoroughly investigated in the literature \cite{walter1965traveling}. Similarly to the DM case, such antennas radiate only at the presence of non-uniformities, curvatures and discontinuities. The CM radiated power of a finite length TL is \cite[Ch. 10]{Balanis:2005:ATA:1208379}
\begin{multline}
P_{rad}^{CM}=\frac{\eta_{0}}{4\pi}\left(\left|I_{f}^{CM}\right|^{2}+\left|I_{b}^{CM}\right|^{2}\right)\cdot \\ \cdot \left(1.415+\log\left(\frac{kL}{\pi}\right)-C_{i}\left(2kL\right)+\sinc\left(2kL\right)\right)
\end{multline}
and the equivalent radiation resistance $R^{CM}$ is
\begin{multline}
R^{CM}=\frac{2P_{rad}^{CM}}{\left(\left|I_{f}^{CM}\right|^{2}+\left|I_{b}^{CM}\right|^{2}\right)}=\\
=\frac{\eta_{0}}{2\pi}\left(1.415+\log\left(\frac{kL}{\pi}\right)-C_{i}\left(2kL\right)+\sinc\left(2kL\right)\right).\label{eq:rcm}
\end{multline}
In order to understand the magnitudes of the radiation resistances in \eqref{eq:rdm} and \eqref{eq:rcm}, we simulated both the equivalent ohmic $R^{Ohm}$ and radiation resistances $R^{CM}$, $R^{DM}$ of a TL made of copper wires with $a$=0.002 m for different cable lengths and frequencies.
\begin{figure}
\subfloat[freq = 100 MHz]{\includegraphics[width=0.5\columnwidth]{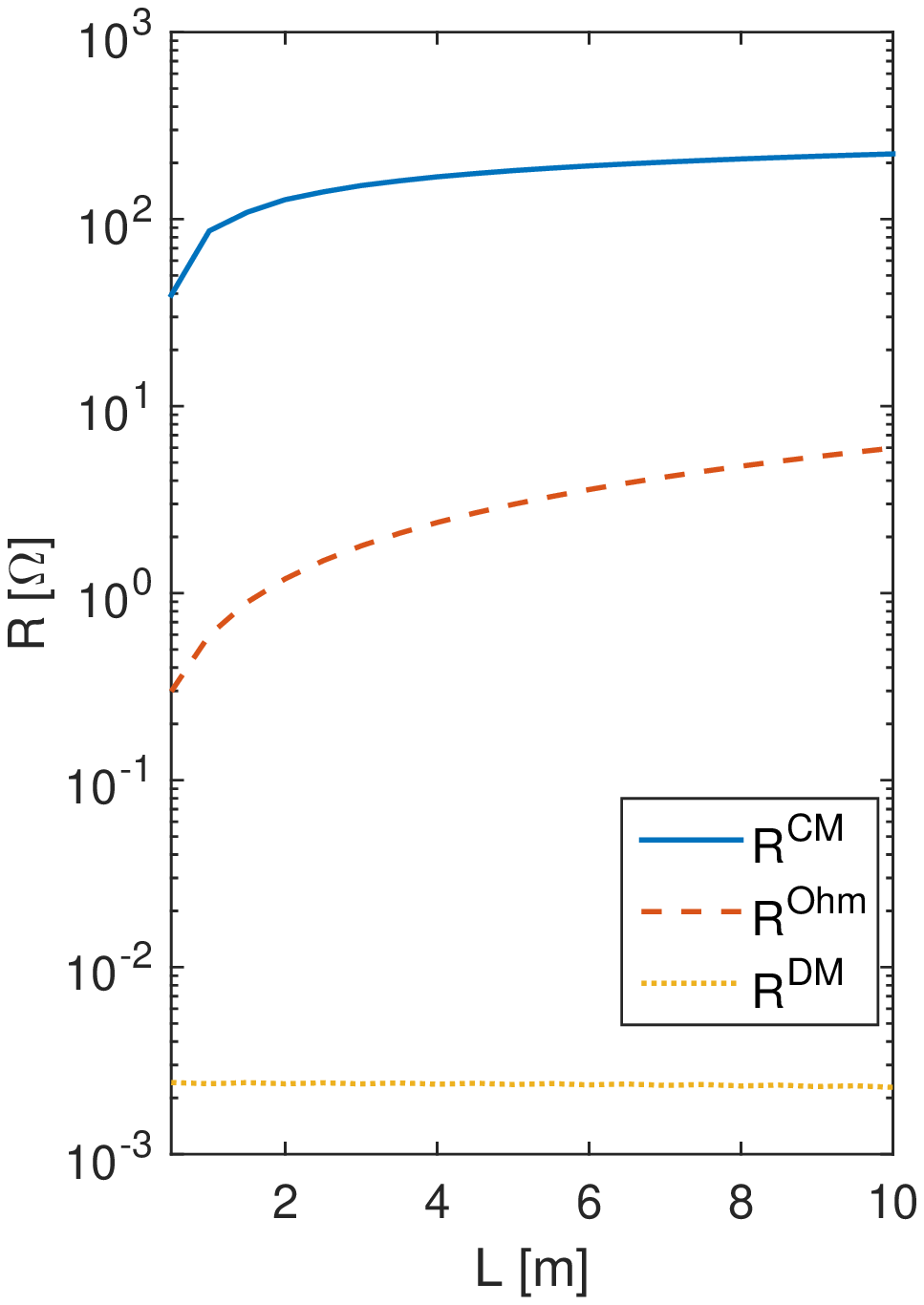}}
\subfloat[L = 1 m]{\includegraphics[width=0.5\columnwidth]{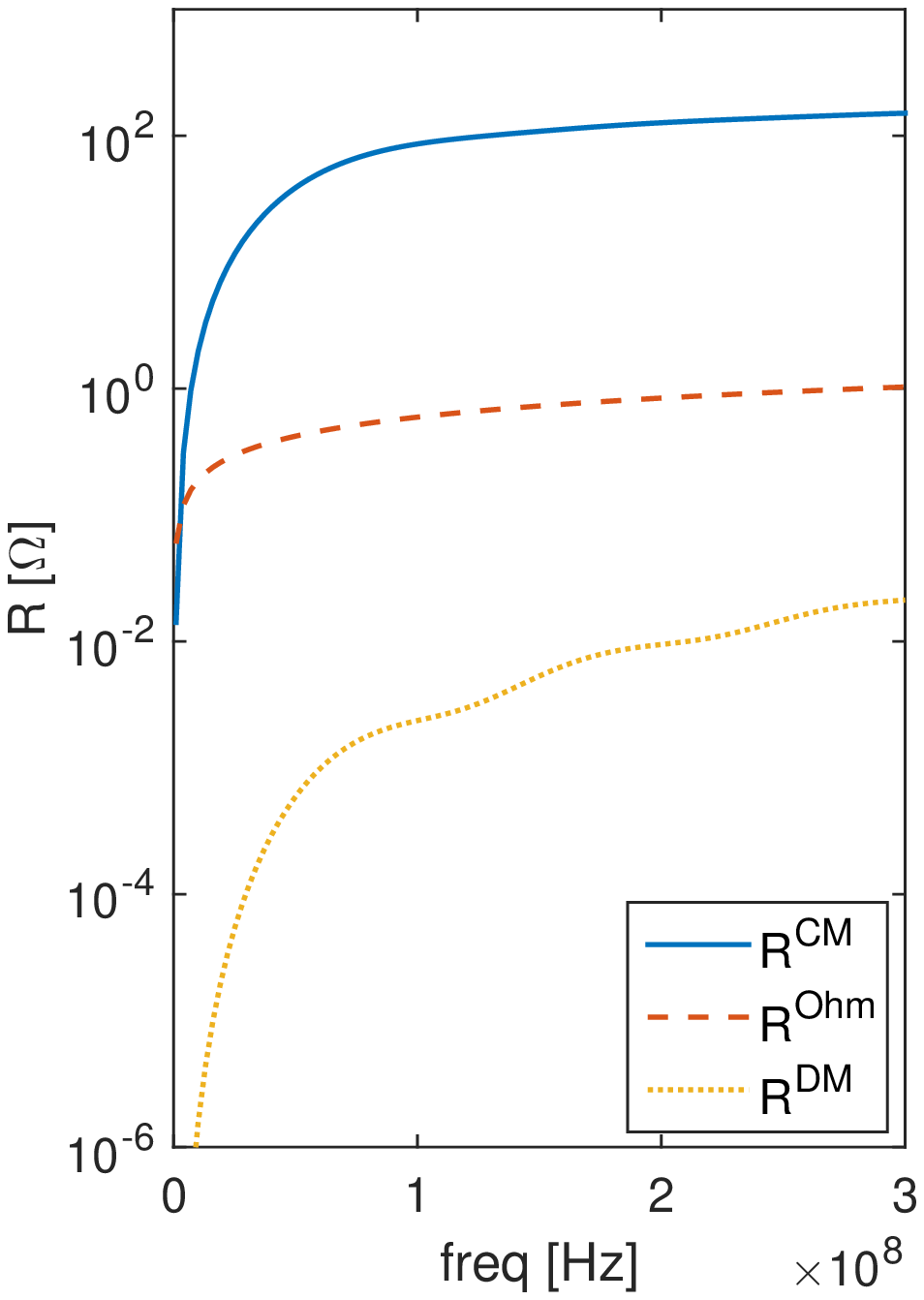}}
\caption{Equivalent ohmic and radiation resistances for different frequencies and line lengths.\label{fig:Equivalent-ohmic-and}}
\end{figure}
It is evident from the plots in \Cref{fig:Equivalent-ohmic-and} that $R^{CM}$ is some order of magnitude greater than $R^{DM}$, while this last one is also a couple of orders of magnitude smaller than $R^{Ohm}$. 
As for the radiation pattern of the power line cables, it is influenced by the load at the end of the line. When we consider a PLN, the load can be a line termination or the equivalent network impedance at a branch node. A matched load causes the propagation of the sole $I_f$ as a traveling wave. In this case most of the radiation is directed to the forward direction, with a main lobe pointing to an angle $\Theta$ over the $I_f$ propagation axis. For longer cables, $\Theta$ and the amplitude of the main lobe tend to decrease and increase respectively. In the case of unmatched loads, both $I_f$ and $I_b$ propagate and their radiation effects sum up. Hence, two main lobes are created in the forward and backward directions, their amplitude being proportional to the square of the forward and reverse currents respectively.

The results presented in this section extend to TL with finite length L what has already been pointed out for infinitesimal dipoles in \cite{7570770}: even CM currents with much lower magnitude than DM currents can produce a similar or even greater radiation. Moreover, any CM current traveling along a cable is deeply attenuated within few meters and its power is radiated instead of being conducted. Hence, it is important in PLC to study how both CM and DM propagate throughout the network, and how different mechanisms lead to CM-DM or DM-CM conversion, which eventually deeply influence the attenuation and the distortion of the communication signals.

\subsection{Radiation model for a two-wire transmission line}
\label{sec:mtrl}
Following the results of the previous section, we herein extend the classical MTL PUL equivalent circuit \cite{Paul:2007:AMT:1554645} for a 3-wire cable to include also the equivalent radiation resistances.

 \begin{figure}
 	\centering 
 	\includegraphics[width=0.5 \textwidth]{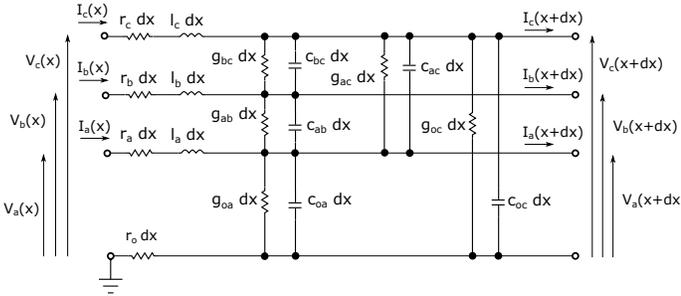}
 	\caption{Per-unit-length MTRL equivalent circuit of a three conductor line over a ground reference.}
 	\label{fig:MTL_circuit_model}
 \end{figure}

\Cref{fig:MTL_circuit_model} shows a line section of four conductors of length $dx$, where $r_i, l_i, c_i, g_i$ denote the PUL resistance, conductance, and capacitance of cable $i$. The mutual inductance, capacitance and conductance take into account the mutual interactions between conductors. 
The bottom conductor is assumed to model a reference conductor, while the other three conductors represent the wires (i.e. Neutral, Phase and Protective Earth) that constitute the power line. We assume the ground, which normally has finite resistance, to be the reference conductor. With this model it is not only possible to analyze the differential transmission between two of the three conductors, but also to analyze the common mode transmission effects on the power line.
 In fact, by considering the MTRL PUL equivalent circuit, the series resistances do not only represent the ohmic losses but also the radiation losses as follows:
\begin{equation}
	r_c = r^{DM}_{c-b}+r^{Ohm}_c,
\end{equation}
\begin{equation}
	r_b = r^{DM}_{b-a}+r^{Ohm}_b,
\end{equation}
\begin{equation}
	r_a = r^{DM}_{a-c}+r^{Ohm}_a,
\end{equation}
\begin{equation}
	r_o = r^{CM}+r^{Ohm}_o,
\end{equation}
where $a$, $b$ and $c$ are the conductors wrapped into the cable and $o$ is the reference ground. With this model, a series resistance for each wire accounts for the radiation losses due to differential signaling between that wire and another one. Moreover, a series resistance on the ground reference accounts for the radiation losses due to the common mode signaling that passes through the cable and closes its path through the ground.

In order for such a model to be extendible to a cable of any length, some simplifications need to be made for the models of $R^{DM}$ and $R^{CM}$, since they do not increase linearly with the cable length. As for $R^{DM}$, we can consider $r^{DM}$ to be equal to $R^{DM}_\infty/\lambda$, where $R^{DM}$ is the total DM radiation resistance of a semi-infinite cable, up to $L=\lambda$. This is because, as shown in Section \ref{sec:radiation}, the value of $R^{DM}$ can be considered a constant function of $L$ for $L>\lambda$. For what matters $R^{CM}$, its trend with the line length is logarithmic. Hence, $r^{CM}$ can be approximated with a piecewise constant value in order to correspond to a $R^{CM}$ with a piecewise linear trend. The number of approximation segments has to be chosen based on a trade-off between the computational complexity and the desired accuracy of the results.

\section{Mode conversion and discontinuities}
\label{sec:network_discontinuities}

In this section, we explain the causes of mode conversion in power line networks, which eventually foster electromagnetic radiation and deteriorate the quality of the signal transmitted through the power line.

The main source of mode conversion in any electric or electronic circuit is system imbalance. Perfectly symmetric cables, circuits and loads would allow an independent transmission of CM and DM signals. While in integrated circuits advanced design strategies permit the physical minimization of mode conversions, few can be done in PLN, since there is no control on the wiring infrastructure, and only the load impedances corresponding to the power line modems can be optimized. Hence, the task of PLC engineers is not much about the minimization of the asymmetries, but it is more about the understanding of mode conversion and the derivation of a phenomenological characterization of it. As a result, new communication algorithms can be developed so that maximum throughput can be achieved yet respecting EMC norms.

We can distinguish between two types of mode conversion that occur in different parts of the network: mode conversion at the interface between the coupler of the power line modem (PLM) and the network, and mode conversion along the network.

\subsection{Coupler-network mode conversion} 
In PLM, unwanted common mode transmission is mostly caused by an unbalance between the conductors on the chip, given by asymmetrical grounding of the traces, or by finite impedance connection between different ground planes \cite{649814}.
Although the design of modern modems allows to produce almost perfectly balanced differential signals, the connection to a long wire antenna like a power line branch causes CM-to-DM conversion. This conversion is due to parasitic coupling between the integrated circuit (IC) and the TL, and to the non-idealities of the coupling transformer. In particular, differential currents in the IC generate magnetic fields that can couple to the TL and generate a CM voltage. Also differential voltages in the IC can generate a CM current on the TL due to capacitive coupling \cite{544310}. 

Similarly, the magnetic field generated by the coupling transformer can impinge upon nearby traces and cables, and the parasitic capacitance between the primary and the secondary acts as a bypass for any CM noise.

\subsection{Conversions along the network}
PLNs are characterized by many sources of unbalance, since many are the devices that constitute the network: power cables, junction boxes, power strips, plugs, sockets and the devices therein branched. Some of the common mechanisms giving rise to mode conversion have already been analyzed in the literature \cite{1629710,Arteche:722106,trcvel14055} and we present them below.

\subsubsection{Conversion to common mode along the line}
A typical power line cable, in both longitudinal and transverse directions, is far from being regular and uniform. Differences in the dielectric material thickness, radius of the copper conductor, relative position of the cables, are usual in this type of network. Also simple cable bends can perturb the symmetry of a line. Each of these discontinuities results in a localized variation of the cable electrical characteristics, which produces mode conversion. 

Moreover, inside junction boxes, power strips, plugs and sockets, the three conductors of a power line are generally unwrapped from the outer sheath. The different conductors are then differently warped and also cut at different lengths before being connected to the junction.

A final source of mode conversion is the difference in ground potentials. Also in indoor environments the ground is characterized by a finite resistance at high frequencies, which fosters the generation and propagation of CM currents.

\subsubsection{Unbalance at the load}
\begin{figure}
 	\centering 
 	\includegraphics[width=0.5 \textwidth]{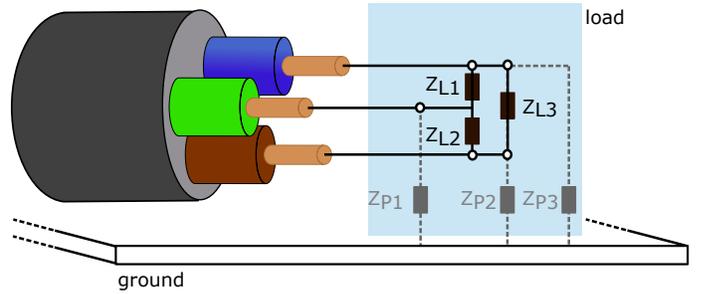}
 	\caption{Schematic mode conversion mechanism at the cable termination.}
 	\label{fig:mode_conversion}
 \end{figure}

Any network load can be represented by the impedances $Z_{L1}$, $Z_{L2}$ and $Z_{L3}$ connected at a line termination, as depicted in \Cref{fig:mode_conversion}. Moreover, the coupling between the lines and the ground is modeled with three parasitic impedances $Z_{P1\dots 3}$. If the impedances of the common mode paths are different $(Z_{P1} \neq Z_{P2} \neq  Z_{P3})$ then part of the DM current will flow through the common mode path. This phenomenon reduces the power available at the load and generates common mode currents. Conversely, the same imbalance also causes CM to generate DM currents, which might corrupt the transmitted DM signal. The imbalance can be caused by different types of couplings between the load and the ground or other objects in the proximity of the load \cite{coupling_EMC}.

\section{Power Line Network Model}
\label{sec:network_model}

Given the MTRL PUL model presented in \Cref{sec:mtrl} and the considerations about mode conversion done in \Cref{sec:network_discontinuities}, we can now introduce a model to study the effect of signal radiation and mode conversion on a generic PLN. First, we rely on the MTL theory to describe the propagation of CM and DM signals along each power line cable. The relations between signals at the two edges of a cable is then described using a mixed-mode scattering matrix (MMSM). MMSMs are also used to characterize all the other devices and loads present in a PLN. By converting the MMSMs to transmission matrices (MMTMs), it is then possible to use the chain rule and represent the whole PLN connecting any two network nodes by a single equivalent MMTM, from which the modal transfer functions can be computed.
We point out that scattering and transmission matrices are here only used for convenience, since high-frequency characterization is usually performed with vector network analyzers that measure scattering parameters. The entire network can be equivalently represented in terms of voltages and currents. The fundamental concept is here to use a mixed-mode representation in order to characterize both DM and CM.

In the following, we summarize the theory about scattering matrices, MMSMs and MMTMs and the conversion equations between each representation. 

\subsection{Three-conductor and earth transmission line equations}

In order to provide a steady-state analysis, we use the phasor representation for the electrical quantities. We denote $V_k(f,x)$ for $k\in \{a,b,c\}$, where $\{a,b,c\}$ are the labels for the three conductors, the voltage phasor associated with the generator circuit and the receiver circuit at a frequency $f$ and coordinate $x$. To simplify the notation, we do not explicitly show the frequency dependency in the following. Solving the Kirchhoff equations for the PUL circuit represented in \Cref{fig:MTL_circuit_model} and letting $dx \rightarrow 0$, provides the telegraph equations \cite{Paul:2007:AMT:1554645}
\begin{align}
\frac{\partial \bm{V}(x)}{\partial x}=&-(\bm{R}+j2\pi f\bm{L})\bm{I}(x),\\
\frac{\partial \bm{I}(x)}{\partial x}=&-(\bm{G}+j2\pi f\bm{C})\bm{V}(x),
\label{eq:telegraph}
\end{align}
where $\bm{V}=(V_a,V_b,V_c)^T$ is the voltage phase vector, $\bm{I}=(I_a,I_b,I_c)^T$ is the current phasor vector, and $\{\dot{}\}^T$ denotes the transposition operator.

Furthermore,
\begin{align}
\bm{R}
=&
\begin{bmatrix}
r_a+r_o & r_o 		& r_o \\
r_o 	& r_b+r_o 	& r_o \\
r_o 	& r_o 		& r_c +r_o \\
\end{bmatrix} \notag\\ 
\bm{L}
=&
\begin{bmatrix}
l_a		& l_{ab} 		& l_{ac} \\
l_{ab} 	& l_b 		& l_{bc} \\
l_{ac} 	& l_{bc} 		& l_c \\
\end{bmatrix} \notag\\
\bm{C}
=&
\begin{bmatrix}
c_{ab}+c_{ac}+c_{ao} 		& -c_{ab} 					& -c_{ac}						 \\
-c_{ab} 					& c_{ab}+c_{bc}+c_{bo} 		& -c_{bc} 							\\
-c_{ac} 					& -c_{bc} 					& c_{bc}+c_{ac}+c_{co} 			\\
\end{bmatrix} \notag\\
\bm{G}
=&
\begin{bmatrix}
g_{ab}+g_{ac}+g_{ao} 	& -g_{ab} 					& -g_{ac}						 \\
-g_{ab} 					& g_{ab}+g_{bc}+g_{bo} 	& -g_{bc} 							\\
-g_{ac} 					& -g_{bc} 					& g_{bc}+g_{ac}+g_{co} 			\\
\end{bmatrix} \notag\\
\label{eq:L_C_R_G_matrices}
\end{align}
are the PUL parameters matrices for the resistance, inductance, capacitance and conductance, respectively. We remark that $\bm{R}$ includes both the ohmic and radiation resistances. Solving \eqref{eq:telegraph} as presented in \cite{versolatto2011an} gives the voltage and current propagation equations

\begin{align}
\bm{I}(x) =& \bm{Z_C^{-1}}\bm{T}(e^{-\bm{\Gamma }x}\bm{V^+}_m - e^{\bm{\Gamma }x}\bm{V^-}_m) \notag\\
\bm{V}(x) =& \bm{T}e^{-\bm{\Gamma }x}\bm{V^+}_m + e^{\bm{\Gamma }x}\bm{V^-}_m ,
\label{eq:V_I_forw_back}
\end{align}	
where $\bm{\Gamma }$ and $\bm{Z_C}$ are the propagation constant and the characteristic impedance of the cable respectively. $\bm{T}$ is a known transformation matrix and $\bm{V^+}_m = \bm{\rho}_L\bm{V^-}_m$, where $\bm{\rho}_L$ is the reflection coefficient given by the load $\bm{Z_L}$ as

\begin{align}
\bm{\rho}_L = \left(\bm{Z_L}-\bm{Z_C}\right)^{-1}\left(\bm{Z_L}+\bm{Z_C}\right).
\label{eq:rho}
\end{align}

\subsection{MTL scattering matrix}
\label{sec:mtlscat}

A power line branch can be in general represented as a multiport system where the signal at N input ports is transferred to N output ports. To derive the scattering matrix of this system, \eqref{eq:V_I_forw_back} has to be solved for $x=0$ and $x=l$, where $l$ is the length of the line, respectively.
The voltages and currents at the input side of the MTL are
\begin{align}
\bm{I}_I =& \bm{Z_C}^{-1}(\bm{V^+}_m - \bm{V^-}_m) \\
\bm{V}_I =& (\bm{V^+}_m + \bm{V^-}_m),
\label{eq:V_I_left_zero}
\end{align}	
while the voltages and currents at the output side of MTL are

\begin{align}
\bm{I}_O =& \bm{Z_C^{-1}}(e^{-\bm{\Gamma }l}\bm{V^+}_m - e^{\bm{\Gamma }l}\bm{V^-}_m) \\
\bm{V}_O =& (e^{-\bm{\Gamma }l}\bm{V^+}_m + e^{\bm{\Gamma }l}\bm{V^-}_m).
\label{eq:V_I_forw_back_right}
\end{align}	
The subscripts $I$ and $O$ denote the input and output ports respectively.
In order to solve these equations, it is necessary to introduce the impedance matrices $Z_I$ and $Z_O$ at the input and the output of the line respectively. These impedance matrices provide a boundary condition through \eqref{eq:rho} to the system and allow it to be solved. 

The resulting scattering matrix $\bm{S}$ is a 2N x 2N matrix, partitioned in 4 NxN matrices as 
\begin{align}
&\bm{S} =
\begin{bmatrix}
\bm{S_{II}} & \bm{S_{IO}}\\
\bm{S_{OI}} & \bm{S_{OO}}\\ 
\end{bmatrix} ,
\label{eq:S_2nx2n}
\end{align}
where $S_{II}$ is  the auto-scattering matrix of the inputs, $S_{OO}$ is  the auto-scattering matrix of the outputs, and $S_{IO}$ and $S_{OI}$ are the transition scattering matrices from the inputs to the outputs and vice versa.
The complete derivation of the scattering matrix \eqref{eq:S_2nx2n} is presented in \cite{kajfez}.

\subsection{Mixed-mode scattering matrix}

In order to improve the understanding of the signal propagation through a transmission line, $\bm{S}$ can be redefined to explicitly show the relation between differential and common modes, using the theory of the mixed-mode scattering matrix (MMSM) \cite{392911}.

The single-ended generic port $j$ is defined by the single-ended voltage and current state vector:
\begin{align}
\bm{s_j} \equiv (V_j,I_j)^T
\label{eq:sing_ended_vi_state_vector}
\end{align}
where the symbol $T$ indicates the transpose operator.

On the other hand, the mixed-mode generic port $jk$ with $(j,k)\in (1\dots M)$, $j\neq k$, where M is the total number of ports, is defined based on the single-ended voltages and currents as:
\begin{align}
V_{jk}^d &\equiv V_j-V_k, \notag\\
I_{jk}^d &\equiv \frac{(I_j-I_k)}{2}, \notag\\
V_{jk}^c &\equiv \frac{(V_j+V_k)}{2},\notag\\
I_{jk}^c &\equiv I_j+I_k  ,
\label{eq:from_single-ended_to_mix_mode_vi}
\end{align}
where the $\{\cdot\}^d$ indicates the differential voltages and currents and the symbol $\{\cdot\}^c$ indicates the common mode ones.
The corresponding mixed-mode voltage and current state vector is:
\begin{align}
\bm{\dot{r_{jk}}} \equiv (V_{jk}^d, I_{jk}^d, V_{jk}^c, I_{jk}^c)^T
\label{eq:mixed-mode_vi_state_vector}
\end{align}
After defining the complex reference impedance as $Z_j$, the forward and reverse pseudowaves for the single-ended ports are defined as:
\begin{align}
&a_j  \equiv \frac{V_j+Ij Z_j}{2 \sqrt{Z_j}}
&b_j  \equiv \frac{V_j-Ij Z_j}{2 \sqrt{Z_j}}
\label{eq:pseudowaves_signle_ended}
\end{align}
A pseudowave state vector of two single-ended ports $j,k$ can be defined as:
\begin{align}
\bm{p_{jk}} \equiv (a_j,b_j,a_k,b_k)^T
\label{eq:sing_ended_ab__jk_state_vector}
\end{align}
For the mixed-mode ports, the reference impedances are defined as $Z{jk}^d \equiv 2\sqrt{Z_j Z_k}$ and $Z{jk}^c \equiv \frac{\sqrt{Z_j Z_k}}{2}$ for the DM and CM respectively. The mixed-mode pseudo waves are therefore defined as:
\begin{align}
&a_{jk}^d \equiv \frac{V_{jk}^d+I_{jk}^d Z_{jk}^d}{2 \sqrt{Z_{jk}^d}} \notag
&b_{jk}^d \equiv \frac{V_{jk}^d-I_{jk}^d Z_{jk}^d}{2 \sqrt{Z_{jk}^d}} \notag\\
&a_{jk}^c \equiv \frac{V_{jk}^c+I_{jk}^c Z_{jk}^c}{2 \sqrt{Z_{jk}^c}}
&b_{jk}^c \equiv \frac{V_{jk}^c-I_{jk}^c Z_{jk}^c}{2 \sqrt{Z_{jk}^c}}
\label{eq:pseudowaves_mixed_mode}
\end{align}
A pseudowave state vector of a single mixed-mode port $jk$ can be defined as:
\begin{align}
\bm{\dot{p_{jk}}} \equiv (a_{jk}^d,b_{jk}^d,a_{jk}^c,b_{jk}^c)^T
\label{eq:mixed-mode_ab__jk_state_vector}
\end{align}
Using \cref{eq:from_single-ended_to_mix_mode_vi,eq:pseudowaves_signle_ended,eq:pseudowaves_mixed_mode}, the relation between single-ended and mixed-mode pseudowaves state vectors becomes 
\begin{align}
\bm{\dot{p_{jk}}} = \Theta \bm{p_{jk}},
\end{align}
where $\Theta$ is reported in \eqref{eq:from_single-ended_to_mixed-mode_pseudowaves}.

\begin{table*}
\centering
\begin{minipage}{\textwidth}
	\begin{align}
	&\bm{\Theta}
	= K
	\begin{bmatrix}
	\sqrt{Z_j}+\sqrt{Z_k}  & -(\sqrt{Z_j}+\sqrt{Z_k}) & \sqrt{Z_j}-\sqrt{Z_k} & -(\sqrt{Z_k}-\sqrt{Z_j})\\
	\sqrt{Z_j}-\sqrt{Z_k} & -(\sqrt{Z_k}-\sqrt{Z_j}) & \sqrt{Z_j}+\sqrt{Z_k}  & -(\sqrt{Z_j}+\sqrt{Z_k})\\
	\sqrt{Z_j}+\sqrt{Z_k}  & \sqrt{Z_j}+\sqrt{Z_k} & \sqrt{Z_j}-\sqrt{Z_k} & \sqrt{Z_k}-\sqrt{Z_j}\\
	\sqrt{Z_j}-\sqrt{Z_k}  & \sqrt{Z_k}-\sqrt{Z_j} & \sqrt{Z_j}+\sqrt{Z_k} & \sqrt{Z_j}+\sqrt{Z_k}\\
	\end{bmatrix}
	&K  = \frac{1}{2 \sqrt{2\sqrt{Z_j Z_k}}}
	\label{eq:from_single-ended_to_mixed-mode_pseudowaves}
	\end{align}
	
	\noindent\makebox[\linewidth]{\rule{\paperwidth}{0.4pt}} 
\end{minipage}
\end{table*}

If $Z_j=Z_k$, then the matrix $\bm{\Theta}$ becomes \cite{1573844}:

\begin{align}
&\bm{\Theta}
= K
\begin{bmatrix}
1  & -1 & 0  & 0 \\
0  & 0  & 1  & -1\\
1  & 1  & 0  & 0 \\
0  & 0  & 1  & 1 \\
\end{bmatrix}
&K  = \frac{1}{\sqrt{2}}.
\label{eq:from_single-ended_to_mixed-mode_pseudowaves_simpified}
\end{align}

The scattering matrix $\bm{S}_{jk}$ for single-ended ports and  $\dot{\bm{S}_{jk}}$ for mixed-mode ports obey the following relations
\begin{align}
&\bm{b}_{jk}\equiv \bm{S}_{jk}\bm{a}_{jk},  &\dot{\bm{b}_{jk}}\equiv \dot{\bm{S}_{jk}} \dot{\bm{a}_{jk}}.
\label{eq:scattering_mx_se}
\end{align}
where $\bm{b}_{jk} =(b_j, b_k)$, $\bm{a}_{jk} =(a_j, a_k)$, $\dot{\bm{b}}_{jk} =(b^d_{jk}, b^c_{jk})$, $\dot{\bm{a}}_{jk} =(a^d_{jk}, a^c_{jk})$.

Reordering the rows of $\bm{\Theta}$, it is possible to compute the relationship between the mixed-mode scattering matrix $\dot{\bm{S_{jk}}}$ and the single-ended scattering matrix $\bm{S}$. To this aim, $\bm{\Theta}$ is split in two rectangular matrices $ \bm{\Theta_1,\Theta_2} $. $\bm{\Theta_1}$ is composed by the first two rows of $\bm{\Theta}$ and $\bm{\Theta_2}$ the last two, respectively. A new matrix $\bm{\Theta_\alpha}$ is made by exchanging the position of $\bm{\Theta_1}$ and $\bm{\Theta_2}$ so that 
\begin{align}
&\bm{\Theta}
=
\begin{bmatrix}
\bm{\Theta_1} \\
\bm{\Theta_2} \\
\end{bmatrix}
&\bm{\Theta_\alpha}
=
\begin{bmatrix}
\bm{\Theta_2} \\
\bm{\Theta_1} \\
\end{bmatrix}.
\label{eq:theta_m}
\end{align}

Finally, from \eqref{eq:theta_m},\eqref{eq:scattering_mx_se}, the direct and inverse transformations linking $\dot{\bm{S}}$ and $\bm{S}$ are:
\begin{align}
&\hat{\bm{S_{jk}}} = {\bm{\Theta_\alpha}}^{-1} \hat{\dot{\bm{S_{jk}}}} \bm{\Theta} \\
&\hat{\dot{\bm{S_{jk}}}} = \bm{\Theta_\alpha} \hat{\bm{S_{jk}}} \bm{\Theta^{-1}},
\label{eq:transform_s}
\end{align}
where $\hat{\bm{S_{jk}}}$ and $\hat{\dot{\bm{S_{jk}}}}$ are defined as

\begin{align}
&\bm{\hat{{S_{jk}}}}
=
\begin{bmatrix}
\bm{S_{jk}} & \bm{I}_2 \\
\bm{I}_2  & \bm{{S_{jk}}^{-1}}\\
\end{bmatrix}
&\dot{\hat{\bm{S_{jk}}}}
=
\begin{bmatrix}
\bm{\dot{S_{jk}}} & \bm{I}_2 \\
\bm{I}_2  & \bm{\dot{{S_{jk}}^{-1}}}\\
\end{bmatrix}
\label{eq:S_explanation}
\end{align}
and $\bm{I}_2$ is the 2x2 identity matrix.

The $\bm{\dot{S_{jk}}}$ matrix is a 2x2 matrix, with the following parameters: $S^{dd}_{jk}$ denoting the differential mode, $S^{cc}_{jk}$ denoting the common mode and $S^{dc}_{jk}$  as well as $S^{cd}_{jk}$ denoting the conversion between both modes.

\begin{align}
&\bm{\dot{S_{jk}}}
=
\begin{bmatrix}
S^{dd}_{jk} & S^{dc}_{jk} \\
S^{cd}_{jk} & S^{cc}_{jk} \\
\end{bmatrix}
\label{eq:S_mix-mode_explanation}
\end{align}

The final scattering matrix respectively for the single-ended and mixed-mode ports are 
\begin{align}
&\bm{b}\equiv \bm{S}\bm{a},  &\dot{\bm{b}}\equiv \dot{\bm{S}} \dot{\bm{a}}.
\label{eq:scattering_mx_tot}
\end{align}
where $\bm{b} =(b_1, \dots ,b_M)$, $\bm{a} =(a_1, \dots ,a_M)$, $\dot{\bm{b}} =(b^d_{jk}, \dots ,b^c_{jk})$ and $\dot{\bm{a}} =(a^d_{jk}, \dots ,a^c_{jk})$. The conversion between $\bm{S}$ and $\dot{\bm{S}}$ is computed by applying \eqref{eq:transform_s} and  to all the possible $j,k$ couples.

\subsection{Mixed-mode transmission matrix}
The description of MTL blocks with the mixed-mode scattering matrix helps to understand the conversion between DM and CM and vice versa. A drowback of the scattering parameters is that they do not allow to easily manipulate the concatenation of multiple scattering matrices. However, MMSMs can be converted to mixed-mode transmission matrices (MMTMs) \cite{1573844}. The system resulting from the connection of $N$ devices can be represented by a MMTM that is given by the multiplication of the MMTMs of the single subsystems. 

\begin{figure}
	\centering 
	\includegraphics[height=0.18 \textheight]{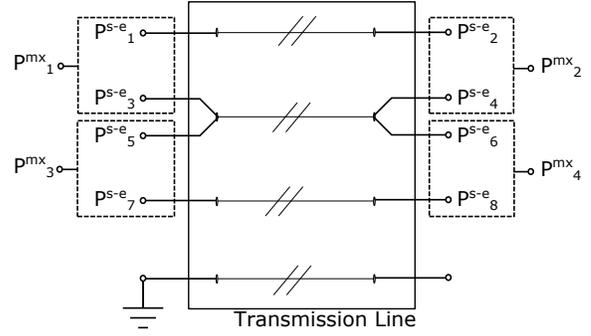}
	\caption{Ports definition: the superscript $\{\cdot\}^{s-e}$ designates single-ended ports; while $\{\cdot\}^{mx}$ designates mixed-mode ports. Each port voltage is referred w.r.t. the ground at the bottom.}
	\label{fig:MTL_to_mixed-mode}
\end{figure}

The conversion procedure comprises: repositioning of the matrix parameters and a linear transformation. The ports should be repositioned in order to reach this form: $\bm{b}=(\bm{b_I},\bm{b_O})^T$ where $\bm{b_I} = (b^d_1,b^c_1,b^d_3,b^c_3)^T$,  $\bm{b_O} = (b^d_2,b^c_2,b^d_4,b^c_4)^T$; $\bm{a}=(\bm{a_I},\bm{a_O})^T$ where $\bm{a_I} = (a^d_1,a^c_1,a^d_3,a^c_3)^T$, $a_O = (a^d_2,a^c_2,a^d_4,a^c_4)^T$, where subscripts $I$ and $O$ refer to the input and output side of the network (see \Cref{fig:MTL_to_mixed-mode}).

The following linear transformation then applies:

\begin{align}
\begin{bmatrix}
\bm{b_R} \\
\bm{a_R} \\
\end{bmatrix}
=
\begin{bmatrix}
\bm{T_{11}} & \bm{T_{12}} \\
\bm{T_{21}} & \bm{T_{22}} \\
\end{bmatrix}
\begin{bmatrix}
\bm{a_L} \\
\bm{b_L} \\
\end{bmatrix}
\label{eq:T_mix-mode_matrix}
\end{align}

Where,
\begin{align}
\bm{T_{11}} =& S_{OI}-S_{OO}S_{IO}^{-1}S_{II}\\
\bm{T_{12}} =& S_{OO}S_{IO}^{-1}\\
\bm{T_{21}} =& -S_{IO}^{-1}S_{II}\\
\bm{T_{22}} =& S_{IO}^{-1}
\label{eq:T_mix-mode_matrix_explaniation}
\end{align}

\section{Application to power line networks}
\label{sec:application}

\begin{figure}
	\centering
		\includegraphics[width=0.40\textwidth]{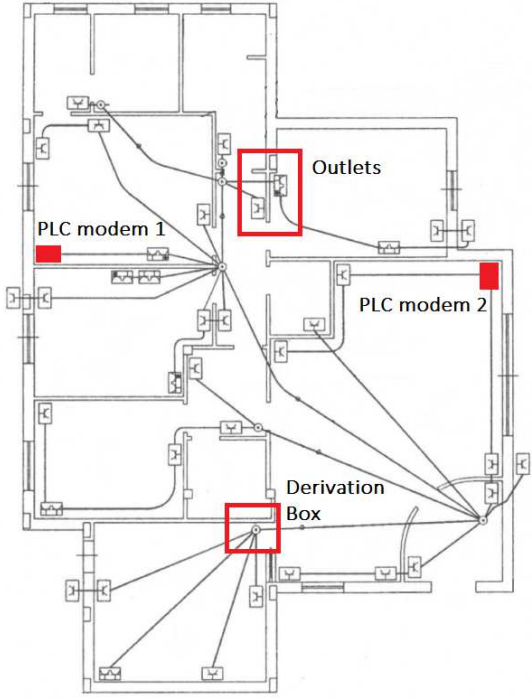}
	\caption{Example of an in-home power line network.}
	\label{fig:inhome_topology}
\end{figure}

Power line networks as the one depicted in \Cref{fig:inhome_topology} are composed by a mixture of different appliances connected by cables that often have different characteristics in different branches \cite{tonello2011bottomup}. Moreover, cable interconnections include junction boxes, power strips, plugs and sockets that increase the network complexity. In order to account for all of these elements when simulating the communication between two PLMs, an equivalent MMTM representation for each element has to be provided. All these matrices are combined using the chain rule \cite{1458855} to compute the total channel frequency response.

\begin{figure}
	\centering 
	\includegraphics[width=0.4 \textwidth]{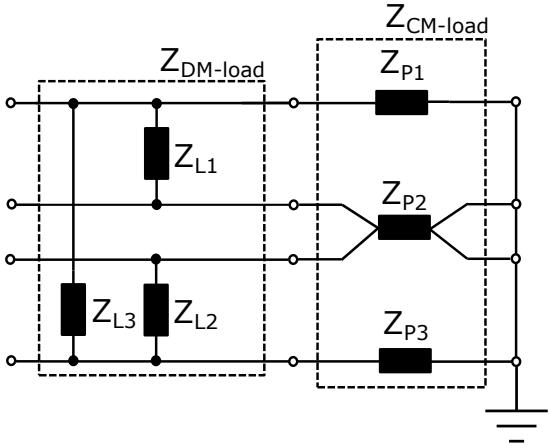}
	\caption{Equivalent model of a generic load.}
	\label{fig:mode_conversion_matrix_anal}
\end{figure}

As for the cables, we represent them using the results reported in \Cref{sec:mtlscat}. We decided to use four input and four output single-ended ports to describe each MTRL block. This configuration is converted in two input and two output mixed-mode ports (see \Cref{fig:MTL_to_mixed-mode}). The defined number of ports has been chosen to obtain a squared shape for the final mixed-mode matrix. The square matrix is easily manipulated for the final conversion to the transmission mixed-mode scattering matrix. Moreover, MIMO PLMs usually communicate through two transmit ports, since only two linearly independent voltage differences can exist between the three conductors of a power line cable. Hence, it make sense to consider the whole PLN to be composed by two input and two output elements.

Using the same approach, each network load can be also represented with a 4-port configuration, as depicted in \Cref{fig:mode_conversion_matrix_anal}, which allows a separate representation of the DM and CM impedances. We point out that $Z_{DM-load}$ in \Cref{sec:mtlscat} is different for different type of loads. In fact, while three phase loads are entirely characterized by $Z_{DM-load}$, single phase loads can be simply described by one of the three impedances $Z_{L_{1\cdots 3}}$, while the other two account for parasitic effects. Finally, PLMs impose two DM impedances on the network termination. The $\bm{S}$ matrix of each load can be in general computed referring to \Cref{fig:mode_conversion_matrix_anal}. In this paper, we report the result obtained for a single phase load characterized by $Z_{L_{1}}$. The scattering matrix referred to $Z_{L_{1}}$ is

\begin{align}
&\bm{S}_{L_{1}} =  \notag \\
&\begin{bmatrix}
\alpha  							& (1+\gamma)\sqrt{\frac{Z_2}{Z_1}}	& \epsilon (1+\delta)				& \epsilon (1+\eta)					\\
(1+\alpha)\sqrt{\frac{Z_1}{Z_2}}	& \gamma 							& \epsilon (1+\delta)				& \epsilon (1+\eta)					\\
\beta (1+\alpha) 					& \beta (1+\gamma) 					& \delta							& (1+\eta)\sqrt{\frac{Z_4}{Z_3}}	\\
\beta (1+\alpha) 					& \beta (1+\gamma)					& (1+\delta)\sqrt{\frac{Z_3}{Z_4}}	& \eta								\\
\end{bmatrix},
\label{eq:differential_load_scattering_matrix}
\end{align}
where
\begin{align}
\alpha &= \frac{(Z_{L_{1}}+Z_3\parallel Z_4)\parallel Z_2 -Z_1}{(Z_{L_{1}}+Z_3\parallel Z_4)\parallel Z_2 +Z_1}\\
\beta  &= \frac{Z_3\parallel Z_4}{Z_3\parallel Z_4 +Z_{L_{1}}}\\
\gamma &= \frac{(Z_{L_{1}}+Z_3\parallel Z_4)\parallel Z_1 -Z_2}{(Z_{L_{1}}+Z_3\parallel Z_4)\parallel Z_1 +Z_2}\\
\delta &= \frac{(Z_{L_{1}}+Z_1\parallel Z_2)\parallel Z_4 -Z_3}{(Z_{L_{1}}+Z_1\parallel Z_2)\parallel Z_4 +Z_3}\\
\epsilon  &= \frac{Z_1\parallel Z_2}{Z_1\parallel Z_2 +Z_{L_{1}}}\\
\eta   &= \frac{(Z_{L_{1}}+Z_1\parallel Z_2)\parallel Z_3 -Z_4}{(Z_{L_{1}}+Z_1\parallel Z_2)\parallel Z_3 +Z_4}
\label{eq:differential_load_scattering_matrix_parameters}
\end{align}
and $Z_{1\dots 4}$ are the reference impedances at port 1\dots 4 respectively and $\parallel$ is the parallel resistance operator.

\section{Measurements results}
\label{sec:comparison_simulation}

In this section, we present measurement results used to justify the need of a mixed-mode cable representation in PLNs.

Our setup is made by a three wire 10 m long power line cable Artic Grade 3183YAG 3x1.5 mm. After measuring its MMSM using a vector network analyzer, we terminated the cable with a load between conductors 1 and 2 having $Z_{L1}=100 \Omega$ and different values of $Z_{P1}$ and $Z_{P2}$. By cascading the cable and load MMSMs, we obtained an equivalent MMSM that has been analyzed as reported in the following.

Assuming to inject only a differential signal between conductors 1 and 2 (corresponding to the mixed-mode port $P_1^{mx}$, see \Cref{fig:MTL_to_mixed-mode}), the ratio between the input and output differential power can be computed as
\begin{align}
r_{d_{1-2}} =& \left|\frac{\text{DM Power at } P_2^{mx}}{\text{DM Power at } P_1^{mx}}\right|,
\label{eq:power_diff_mx}
\end{align}
while the ratio between the input power and the common mode output power is:
\begin{align}
r_{dc_{1-2}} =& \left|\frac{\text{CM Power at } P_2^{mx}}{\text{DM Power at } P_1^{mx}}\right|.
\label{eq:power_diff_cm_mx}
\end{align}
The complex power $P_n+iQ_n$ at the port $n$ either for the DM or the CM is computed using the relation
\begin{align}
	P_n+iQ_n =& \left(a_n+b_n\right)\left(a_n-b_n\right)^*,
\end{align}
where $\{\cdot\}^*$ is the complex conjugate operator.

\begin{figure}
	\centering
		\includegraphics[width=0.50\textwidth]{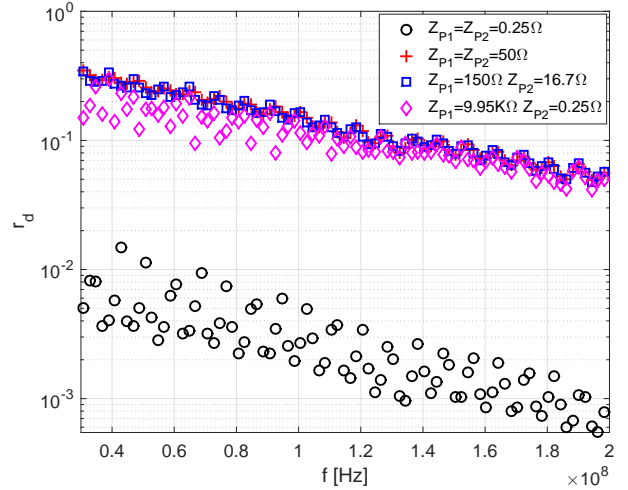}
		\caption{$r_{d}$ as function of the frequency for different CM loads. The DM load is fixed at 100 $\Omega$.}
	\label{fig:10m_unroll_power_ratio_T_rd}
\end{figure}

\Cref{fig:10m_unroll_power_ratio_T_rd} depicts the frequency trend of $r_{d_{1-2}}$ for different values of $Z_{P1}$ and $Z_{P2}$. Circles and plus symbols are used to represent the results for balanced loading with 0.25 $\Omega$ and 50 $\Omega$ respectively. Square symbols represent the results for $Z_{P1}=$ 150 $\Omega$ and $Z_{P2}=$16.7 $\Omega$, while diamond symbols represent the results for $Z_{P1}=$ 9.95 k$\Omega$ and $Z_{P2}=$0.25 $\Omega$.  When $Z_{P1}=Z_{P2}$, the low values of $r_{d_{1-2}}$ are only due to ohmic and radiation losses. The impedance imbalance represented by the square symbols does almost not affect the DM transmission, while the reduction of the value of $r_{d_{1-2}}$ becomes evident only for extreme imbalances (diamond symbols). Finally, when both $Z_{P1}$ and $Z_{P2}$ have very low values (circles), DM transmission is significantly reduced.

\begin{figure}
	\centering
		\includegraphics[width=0.50\textwidth]{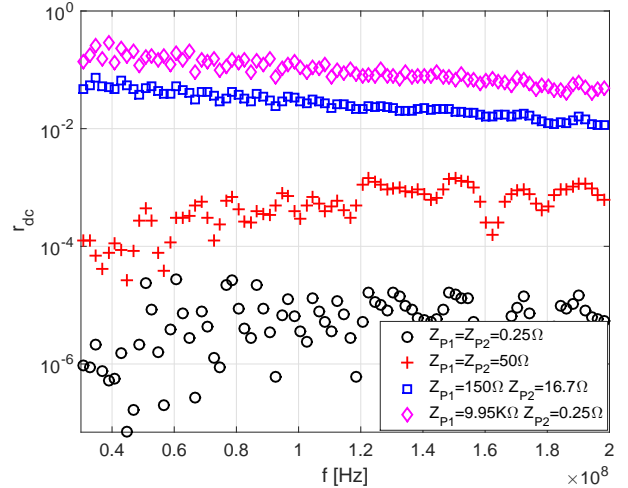}
	\caption{$r_{dc}$ as function of the frequency for different CM loads. The DM load is fixed at 100 $\Omega$.}
	\label{fig:mx10m_power_ratio_T_rdc}
\end{figure}

\Cref{fig:mx10m_power_ratio_T_rdc} depicts the frequency trend of $r_{dc_{1-2}}$ for different values of $Z_{P1}$ and $Z_{P2}$. While the two cases with $Z_{P1}=Z_{P2}$ exhibit very low values of $r_{dc_{1-2}}$, the mode conversion is significantly higher for $Z_{P1}\neq Z_{P2}$. This confirms that, when the CM load is balanced, the measured mode conversion is due to little imbalances internal to the cable, while when the CM load is unbalanced, the load asymmetry effect dominates on the overall mode conversion. Moreover, the greater the load imbalance is, the greater is the power transfer from differential to common mode. In fact, a comparison between \Cref{fig:10m_unroll_power_ratio_T_rd} and \Cref{fig:mx10m_power_ratio_T_rdc} shows that the power converted to CM when $Z_{P1}=$ 150 $\Omega$ and $Z_{P2}=$16.7 $\Omega$ (square symbols) is about one order of magnitude less than the transmitted differential power. Conversely, when $Z_{P1}=$ 9.95 k$\Omega$ and $Z_{P2}=$0.25 $\Omega$ (diamond symbols) so that the CM loads are more unbalanced, the transferred power is almost equally split between differential and common mode.

\section{Conclusions}
\label{sec:conclusions}

In this paper, we presented a novel power line network model that takes into account the effect of EMI and also mode conversions occurring along the medium on the signal transfer function. We firstly showed that radiation in PLC  occurs at the network discontinuities and its radiation pattern depends on the length of each branch. The electro magnetic radiation of the cables is accounted in the PUL equivalent circuit of the MTRL by a series resistance for each wire. 
Since the radiation due to CM signals is much prominent than radiation due to DM signals, we proposed to describe each element of a PLN with its equivalent MMSM, which accounts for mode conversions. By cascading the MMTM blocks corespondent to each cable section and each device/load, the overall PLN connecting any two nodes can be represented by an equivalent MMTM. We showed that such matrix can be used to compute the overall mode conversion of the network and we tested our model measuring the mode conversion of a loaded cable. The results of this paper open a path for future research endeavors that will be directed towards the development of a comprehensive power line network simulator that incorporates conducted and radiated phenomena.

\bibliographystyle{IEEEtran}
\bibliography{IEEEabrv,biblio}

\end{document}